\documentclass[twocolumn,showpacs,preprintnumbers,amsmath,amssymb,eps]{revtex4-1}

\usepackage{graphicx}
\usepackage{dcolumn}
\usepackage{bm}
\usepackage{color}

\begin{document}


\title{
Theory of the Fermi Arcs, the Pseudogap, $T_c$ and the Anisotropy in k-space of Cuprate Superconductors
}
\author{E. V. L de Mello }
\affiliation{%
Instituto de F\'{\i}sica, Universidade Federal Fluminense, Niter\'oi, RJ 24210-340, Brazil\\}%

\date{\today}

\begin{abstract}
The appearance of the Fermi arcs or gapless regions at the nodes
of the Fermi surface just above the critical temperature is
described through self-consistent calculations in an electronic 
disordered medium. We develop 
a model for cuprate superconductors based on an array of
Josephson junctions formed by grains of inhomogeneous
electronic density derived from a phase separation transition. 
This approach provides physical
insights to the most important properties of these materials 
like the pseudogap phase as forming by the onset of local (intragrain) 
superconducting amplitudes and the zero resistivity
critical temperature  $T_c$ due to  phase coherence 
activated by Josephson coupling. The formation of the Fermi arcs
and the dichotomy in k-space follows from the direction dependence
of the junctions tunneling current on the d-wave symmetry
on the $CuO_2$ planes. We show that this semi-phenomenological approach 
reproduces also the main future of the cuprates phase diagram.

\end{abstract}


\maketitle

A convincing explanation to the origin of the pseudogap and 
the dome-like shape of the superconducting
critical temperature $T_c(p)$ on the doping level $p$ 
of the copper oxide superconductors has
become a major long standing challenge in condensed matter physics.
The lack of a well accepted theory is likely to be due to the nanoscale
complexity and the intrinsically inhomogeneous electronic 
structures\cite{Muller,BianconiB} that are not easy to be 
mathematically characterized. Recently, 
new techniques were developed to 
control the level of phase separation of dopant 
oxygen interstitials in $La_2CuO_{4+y}$ and established a direct 
correspondence between the degree of order and the values of
$T_c$\cite{Fratini,Poccia} paving the way to new theoretical
treatments.

Electronic inhomogeneities in
cuprates were verified by several different experiments: neutron
diffraction\cite{Tranquada,Bianconi,Bozin}, muon spin relaxation
($\mu SR$)\cite{Uemura,Sonier}, nuclear quadrupole resonance (NQR) 
and nuclear magnetic resonance (NMR)\cite{Singer,Keren}
have  detected some form of disordered local electronic densities. 
The nanometer spatial
variations of the  electronic gap amplitude $\Delta(\vec r)$
measured by atomically resolved spectroscopy such as scanning
tunneling microscopy (STM)\cite{McElroy,Gomes,Pasupathy,Pushp} is
possibly related with the charge inhomogeneities. Angle resolved
photon emission spectroscopy (ARPES) found a large anisotropy in
k-space: a larger gap at the leading edge of the Fermi surface 
or antinodal (along the $Cu-O$ bonds) direction (($\pm \pi,0$) and 
($0,\pm\pi$))\cite{Damascelli} that remains well above $T_c$ 
and a small d-wave like gap that vanishes at the nodal direction. 

More recently, ARPES experiments measured nodal gapless Fermi arcs 
starting at $T_c$ and increasing with the temperature while 
the  antinodal gaps remained almost constants\cite{Lee}. 
Another group revealed a metallic dispersion along the Fermi arcs and
a Bogoliubov-like dispersion toward the antinodal regions either
above and below $T_c$\cite{Kanigel} demonstrating the presence 
of the superconducting amplitudes above $T_c$. Electronic 
Raman scattering experiments\cite{LeTacon} showed that the 
nodal gap is connected with $T_c$ and the antinodal with the 
pseudogap. Intrinsic tunneling spectroscopy also measured a
gap that closes at $T_c$ and a pseudogap that virtually does
not change with the temperature and remains above $T_c$\cite{Krasnov},
in agreement with two energy scales in cuprate superconductors\cite{Lee,LeTacon}.

To describe these complex phenomena some theories producing 
phase separation have been suggested, mainly based on doped Mott-Hubbard
insulators\cite{Zaanen,Schulz,Emery,Kivelson,Gorkov1}. While 
they describe some of the observed features, 
they fail to predict all the details related with 
real space inhomogeneities and the k-space dichotomy.
While these theories
describe some of the observed features of cuprates, 
they fail, for instance, to predict all the details related with 
real space inhomogeneities and the k-space dichotomy. To deal with
this problems some theories considered the presence of impurities 
or inhomogeneities in the local electronic distribution  as a starting 
point to derive the superconducting properties\cite{OWK,Mello03}.
Some others approaches used the influence of a mesoscopic phase separation 
in the appearance of superconductivity to derive the general properties 
of cuprates\cite{Yukalov,Mello04}.

In this letter we show that it is possible to find an unified
explanation to all these experiments starting with an electronic 
phase separation EPS forming small granular regions where 
isolated superconducting amplitudes may be formed. The origin of 
the EPS may be due to the lower free energy of the low density 
anti-ferromagnetic domains\cite{Mello09}.
These local d-wave superconducting amplitudes in these
islands form an array of multiple Josephson 
junctions\cite{PhysicaC2011,prl2011}. 
In a typical d-wave superconductors junction the 
direction dependence of the tunnel matrix elements that 
describe the barrier is relevant\cite{Bruder}. In the case of
cuprates, all the domains have d-wave pair wave functions with 
the same direction with respect to the $a$ and $b$ crystal axis. 
Phase fluctuation above $T_c$ generates the gapless region 
that increases with temperatures at 
the nodal direction where the d-wave superconducting gap is smaller
and accounts for the k-space anisotropy .


To describe mathematically the electronic phase separation of high
$T_c$ oxides we use the time 
dependent Cahn-Hilliard (CH) equation\cite{CH}.
In this approach it is possible to follow
the formation of domains with distinct local densities below 
the phase separation transition temperature $T_{PS}$.  
$T_{PS}=T_{PS}(p)$ is likely to be close the upper pseudogap 
temperature since $T_{PS}(p)$ may be the cause of some observed
high temperature anomalies\cite{TS}.
The time $t$ in the CH equation is related with the temperature $T$;
larger times correspond to lower temperatures below $T_{PS}(p)$ and
consequently, larger disorder\cite{Otton,Mello04,Mello09,Mello11}. In 
the disordered phase the appropriate order parameter is the 
difference between the local and the average charge density $p$,
$u(p,i,T)\equiv (p(i,T)-p)/p$. $u(i,T)=0$ corresponds to 
the homogeneous case above $T_{PS}(p)$ and the Ginzburg-Landau 
(GL) free energy functional  is given by the usual
power expansion,

\begin{eqnarray}
f(u)= {{{1\over2}\varepsilon^2 |\nabla u|^2 +V_{GL}(u,T)}}.
\label{FE}
\end{eqnarray}
Where the potential ${\it V_{GL}}(u,T)= -A^2(T)u^2/2+B^2u^4/4+...$,
$A^2(T)/B=\alpha(T_{PS}(p)-T)$ for $T<T_{PS}$, $\alpha$ is a constant. 
$\varepsilon$ gives the
size of the barrier between the low and high density phases
\cite{Otton,Mello04}. The CH equation can be written\cite{Bray} 
in the form of a continuity equation of the local density of free energy $f$,
$\partial_tu=-{\bf \nabla.J}$, with the current ${\bf J}=M{\bf
\nabla}(\delta f/ \delta u)$, where $M$ is the mobility or the
charge transport coefficient that sets the phase 
separation time scale. Therefore,
\begin{eqnarray}
\frac{\partial u}{\partial t} = -M\nabla^2(\varepsilon^2\nabla^2u
- A^2(T)u+B^2u^3).
\label{CH}
\end{eqnarray}

\begin{figure}[ht]
    \begin{center}
     \centerline{\includegraphics[width=7.0cm]{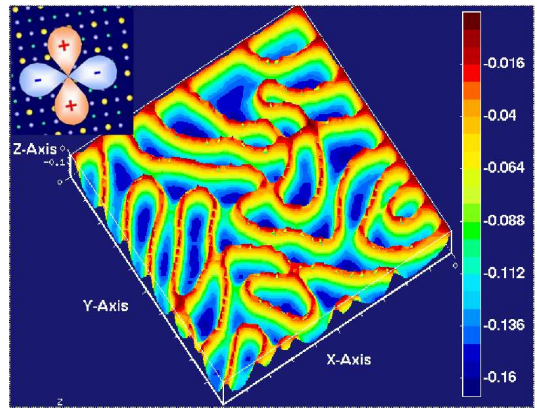}}
\caption{ (color online) Typical solution of the free energy
potential ${\it V_{GL}}(u,T)$ on a $50 \times 50$ lattice. At low 
temperatures the charges are attracted to the grains, similar to
a granular superconductor. On the top left is shown the d-wave 
superconducting amplitude in these bound regions.
}
\label{MapVu} 
\end{center}
\end{figure}

In Fig.(\ref{MapVu}), we show a typical ${\it V_{GL}}(u,T)$
simulation where the barriers between different grains are clearly visible. 
We have argued in previous works\cite{PhysicaC2011,prl2011} that the particles
may become confined in these domains increasing the probability
of Cooper pair formation. In our calculations we take into account that
these free energy barriers are essentially constant at
low temperatures and varies as $(T_{ps}-T)^{1.5}$ near the 
transition as shown by CH\cite{CH}.
                                                                   
%
Then, after performing the  CH simulations, 
the local disorder density $p(i,T)$ is used as
the {\it initial input} and {\it it is  maintained fixed} throughout the
self-consistent Bogoliubov-deGennes (BdG) method. This method
yields local varying d-wave amplitudes $\Delta_c(i,p,T)$ as
demonstrated previously\cite{PhysicaC2011,prl2011,Mello09,Mello11}.

\begin{figure}[ht]
     \centerline{\includegraphics[width=6.0cm,angle=-90]{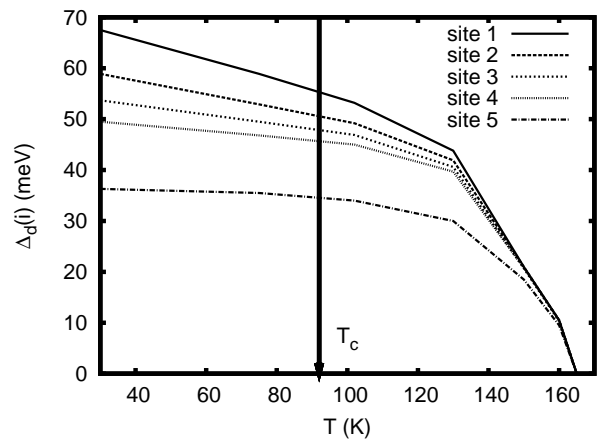}}
\caption{The temperature evolution of $\Delta_d(i)$ at five different locations 
of a $p=0.15$ compound. The mean-field BdG calculations all vanishes at the same 
temperature $T^*(0.15)\approx 165$K (above $T_c(0.15)=92$K). The low 
temperature values  between 30-70meV are in the energy
range of the Bi2212 LDOS gaps measured by STM.}
\label{DeltaT} 
\end{figure}

As mentioned above, $V_{GL}(p,T)$ favors the charge to be
confined into the  domains acting as a catalyst to 
Cooper pair formation. We have showed that the two-body
attractive potential of the BdG equation is scaled by the height 
of the free energy barriers shown in Fig.(\ref{EJ}). In
Ref.(\cite{prl2011}) we called it the grain boundary potential
$V_{gb}$. Here we generalize it to any doping value by
the expression, $V_{gb}(p)=-14.2+45\times p$ (eV). With this 
two-body attractive potential
a typical BdG solution $\Delta_d(p=0.15,i,T)$ is shown 
in Fig.(\ref{DeltaT}) for five different grains  ``i``.
The low temperature values between 30-70meV are in the energy
range of the LDOS gaps measured by STM 
experiments on Bi2212\cite{McElroy,Gomes,Pasupathy,Pushp}.

To obtain the experimental values of $T_c(p)$, we use an
approach similar to a system composed of granular 
superconductors\cite{Merchant}. Our main proposal is that the
superconducting transition occurs in two steps as T 
decreases\cite{PhysicaC2011,prl2011,Mello11}: first by 
intra-grain superconductivity and than by Josephson coupling
with phase locking at low temperatures. 

These two completely different calculations, yielding two different energy
scales, are motivated by the two
energy scales found in most cuprates mentioned above\cite{Damascelli,LeTacon}, and
also in the two different regimes of the fluctuation magnetoconductivity
in a $YBa_2Cu_3O_7$ single crystal\cite{Pureur}. These systematic measurements
detected an effectively two-dimensional (2D) regime far above $T_c$, 
that can be interpreted in our approach by the isolated
superconducting grains in the $CuO$ without phase 
coherence. Decreasing the temperature towards $T_c$ a crossover to
a three-dimensional (3D) Gaussian regime sets in as the different superconducting
regions or grain become connected and increase coherence. 
Very close to $T_c$ a critical regime characteristic
of a 3D XY university class is measured which is consistent with a granular 
superconductors in which each superconducting grain develops its
own phase $\Phi$ that may oscillate but become locked at $T_c$, yielding long
range order.

\begin{figure}[ht]
     \centerline{\includegraphics[width=8.0cm]{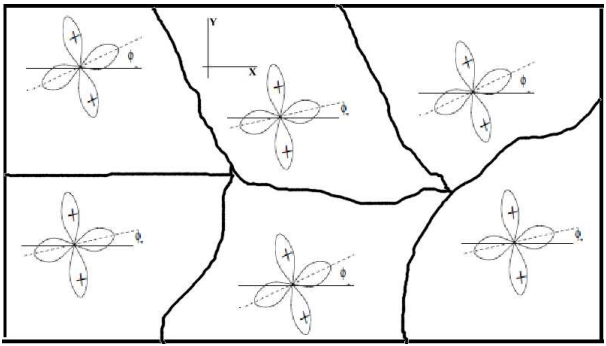}}
\caption{ Schematic figure representing the array of Josephson junctions
on a $CuO_2$ plane.
The local dependent amplitudes $\Delta(p,i,T)$ 
are represented by the orbitals and the phase angles $\phi$ 
oscillates around the crystal $a$-axis direction due to the symmetry of the 
d-wave on the $CuO_2$ plane.
}
\label{JJsDdw} 
\end{figure}
\begin{figure}[ht]
\begin{center}
  \begin{minipage}[b]{.1\textwidth}
    \begin{center}
     \centerline{\includegraphics[width=5.0cm,angle=-90]{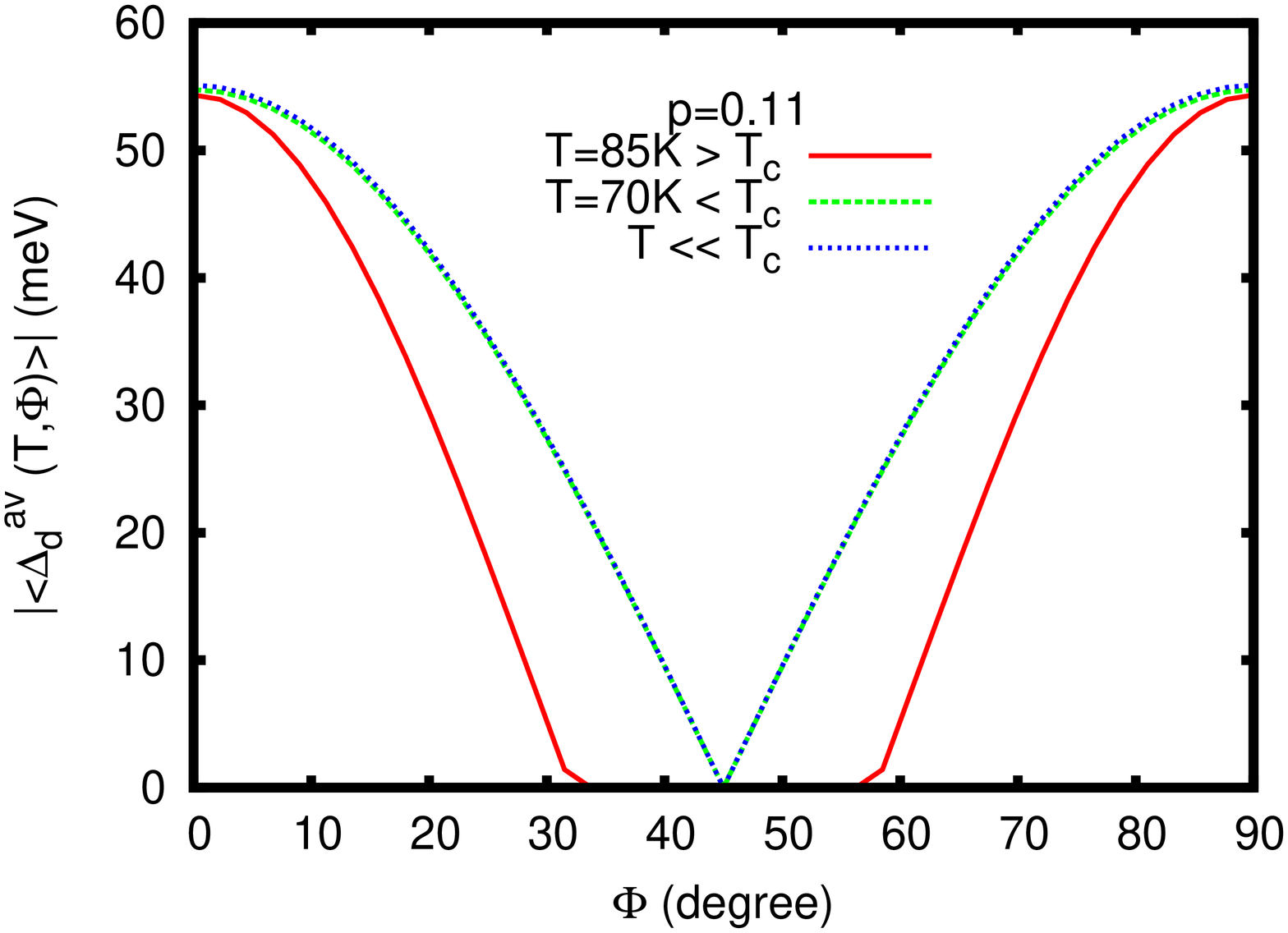}}
     \centerline{\includegraphics[width=5.0cm,angle=-90]{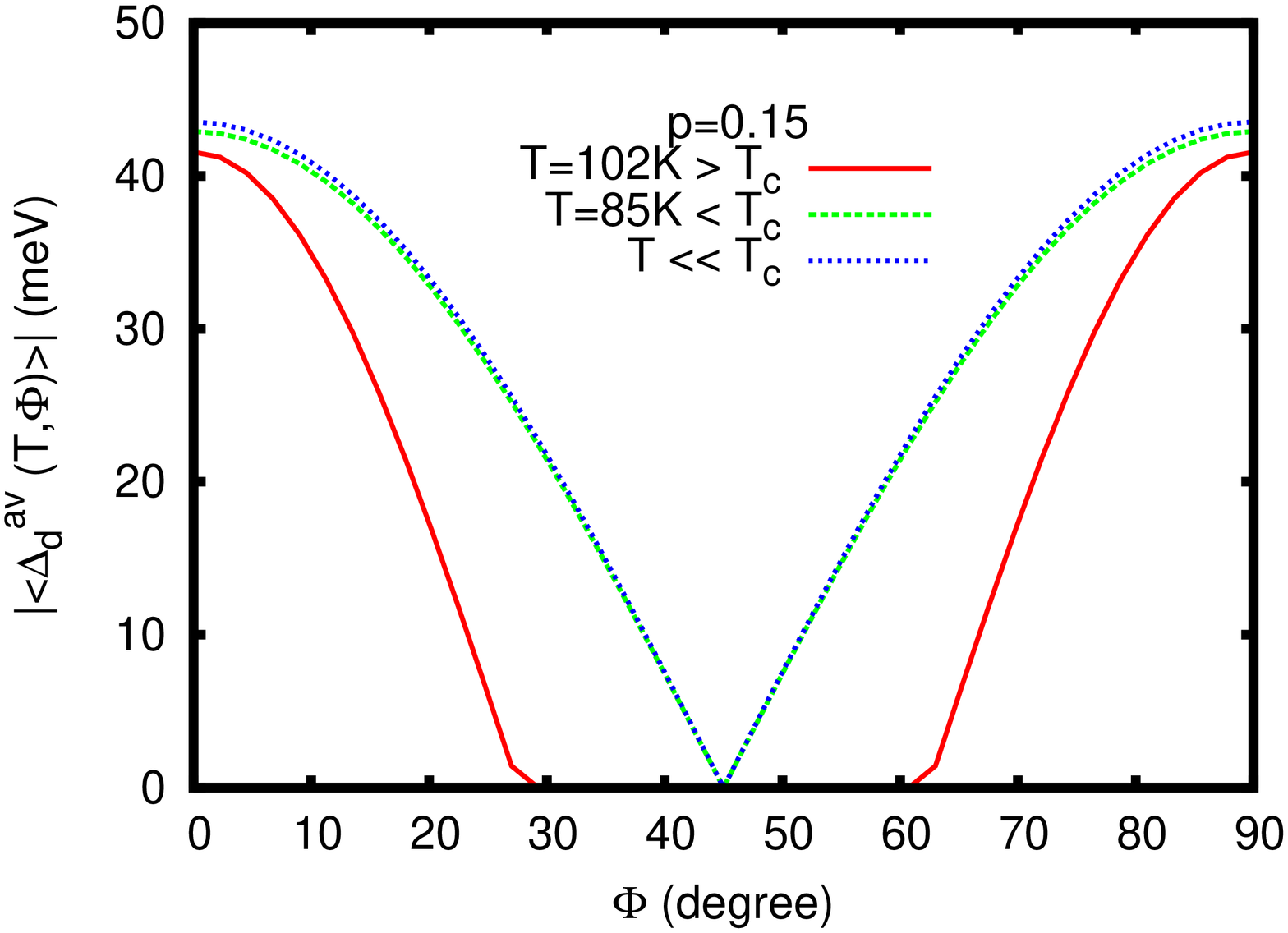}}
    \centerline{\includegraphics[width=5.0cm,angle=-90]{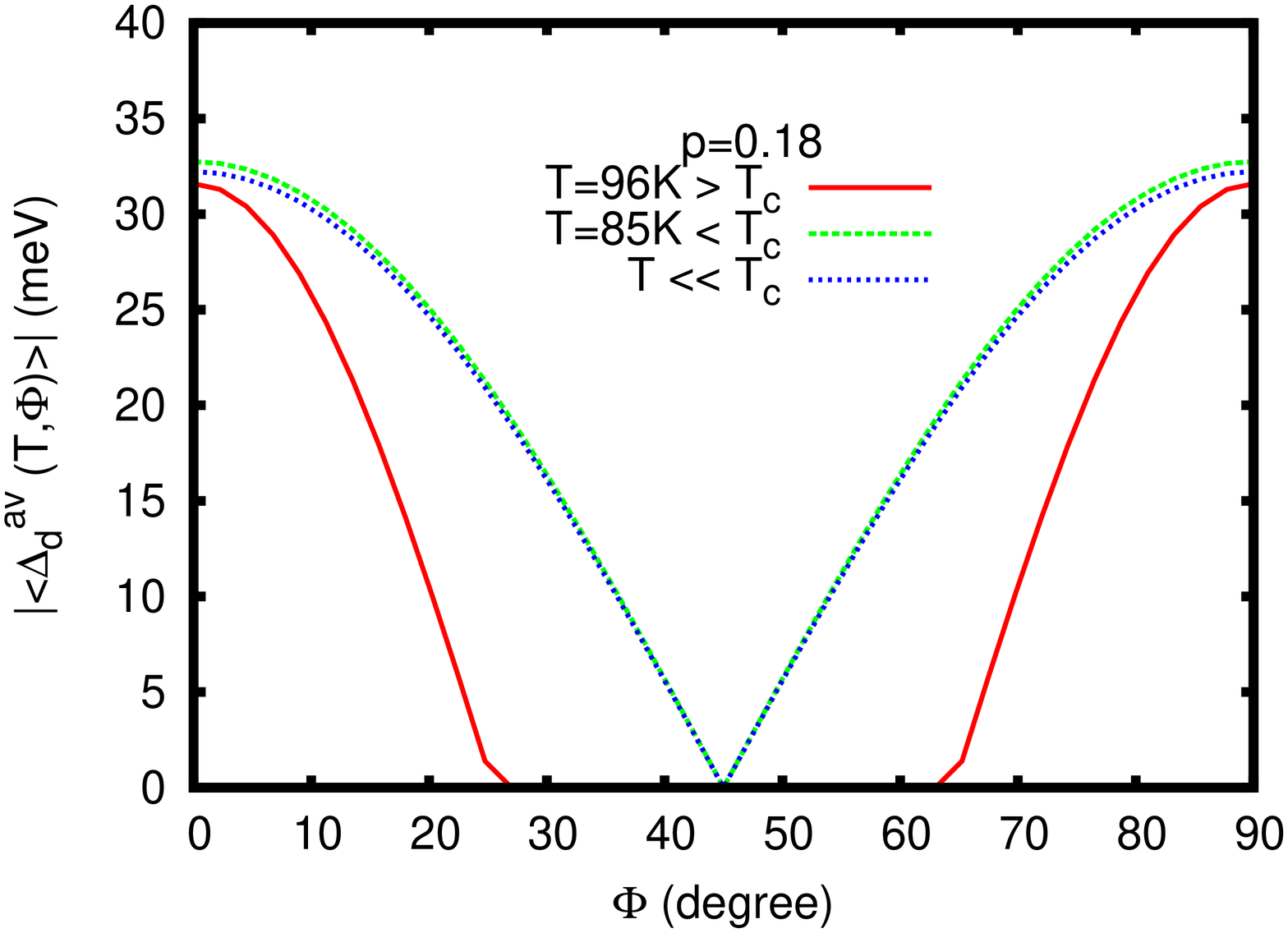}}
    \end{center}
 \end{minipage}
\caption{ The absolute value of the spatial and $\phi$ average, 
$<\Delta_d^{av}(T,\Phi)>$, as function of the k-space polar angle 
$\Phi$. The developing of the Fermi arcs
around the nodal direction above $T_c$ is a consequence of the lack of
phase coherence and, as measured 
by ARPES\cite{Kanigel,Lee}, $<\Delta_d^{av}(T,\Phi)> \le K_BT$.}
\label{DkT}
 \end{center}
\end{figure}

Consequently, the system is regarded as an
array of Josephson junction as shown schematically in Fig.(\ref{JJsDdw}).
The phenomenology of Josephson tunnel junctions between d-wave
superconductors have been developed by Bruder et al\cite{Bruder}.
They calculate the tunnel matrix elements in second-order perturbation
theory for two superconductors (1 and 2) with superconducting amplitude 
$\Delta_{d,1/2}(i,T,\Phi)=\Delta(i,T)cos[2(\Phi-\phi_{1/2})]$.
We use their results to an array of junction as schematically 
shown  in Fig.(\ref{JJsDdw}),
where the local amplitudes $\Delta_d(i,T)$ are 
represented by the size of orbitals and the phase angles $\phi$ have 
been drawn around x-direction. $\Phi$ is the polar angle
of the first Brillouin zone. Due to the thermal energy  
the gap must be averaged over $\phi$, then
$<\Delta(p,T,\Phi)> =(\int^{\Delta\phi} \Delta_d^{av}(p,T)cos[2(\Phi-\phi)]d\phi)/\Delta\phi$.
And  $\Delta^{av}_d(p,T,)\equiv \sum_i^N \Delta_d(p,i,T)/N$ 
is the spatial average superconducting amplitude. 
At $T\le T_c$ the values of $\phi_{1/2}$ are essentially
locked. 
At $T \ge T_c$ the phase angles
decouple and oscillate within a maximum  amplitude 
that increases with $T$ as shown in Fig.(\ref{JJsDdw}).
In this case $<\Delta_d^{av}(p,T)cos[2(\Phi-\phi)]>$
vanishes when  $2(\Phi-\phi)\approx \pm\pi/4$ or it is
less than $K_BT$. The results are
shown in Fig.(\ref{DkT}) and compare well with
the $\Phi$ dependent ARPES measurements of Lee at al\cite{Lee}.
The appearance of a gapless region around the nodal direction ($\Phi=\pm\pi/4$)
above $T_c$ while the antinodal ($\Phi=\pm n\pi/2$, n=1,2,..) gap remains almost 
constant is concordant with the data of Kanigel et al\cite{Kanigel}. 
They found that the electronic
dispersion is either metallic along the Fermi arcs or exhibits 
Bogoliubov-like behavior near the nodes at temperatures above $T_c$.
This behavior can be understood with the results for $\Delta_d(i,p,T) \times T$
shown in Fig.({\ref{DeltaT}) and $\Delta^{av}_d(p,T) \times \Phi $ 
of Fig.({\ref{DkT}).

To calculate the Josephson coupling energy that is connected 
with the onset of phase coherence, we recall the work of 
Bruder et al\cite{Bruder}. They found that the tunneling current 
behaves in a similar fashion of s-wave superconductors junction and the 
leading behavior is determined by tunneling from a gap node in 
one side of a junction into the effective gap in the other side. 
Consequently, as a first approximation to the Josephson coupling energy $E_J$,
we adapt the theory of s-wave granular superconductors\cite{AB} to
the $\Delta^{av}_d(p,T)$ in the grains.
\begin{eqnarray} 
E_J(p,T) = \frac{\pi h\Delta^{av}_d(p,T)}{4 e^2 R_n(p)}
tanh(\frac{\Delta^{av}_d(p,T)}{2K_BT}).
\label{EJ} 
\end{eqnarray} 
Where $R_n(p)$ is the normal resistance of the compound. We use here values
proportional to the planar resistivity $\rho_{ab}$ 
measured on the $La_{2-p}Sr_pCuO_2$ series\cite{Takagi} and the estimated 
values of $R_n(p)e^2/h$ are plotted in Fig.(\ref{DoRnEJTc}).
Since the onset of phase coherence occurs when the thermal energy 
equals the Josephson coupling, $K_BT(p)=E_J(p,T)$,
we obtain one of our most important result; the derivation of the
dome-like shape dependence of $T_c(p)=E_J(p,T_c)/K_B$, as plotted  
in Fig.(\ref{DoRnEJTc}).

\begin{figure}[ht]
\begin{center}
     \centerline{\includegraphics[width=5.5cm,angle=-90]{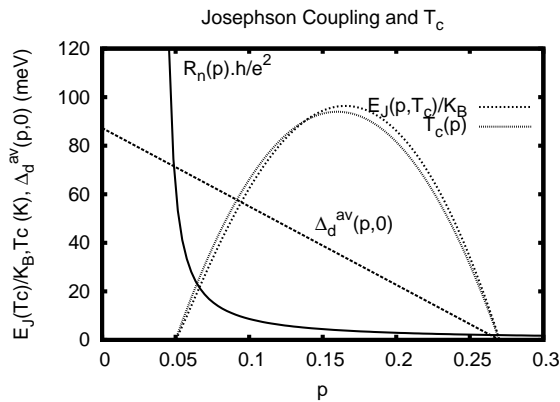}}
\caption{ The temperatures which $E_J(p,T)=K_BT(p)$ yield 
$T_c(p)=E_J(p,T_c)/K_B$, for comparison we plot also  the 
formula $T_c(p)=95(1-82.6(p-0.16)^2)$. The experimental resistivity  $R_n(p)e^2/h$
used in $E_J(p,T)$ (Eq.\ref{EJ}) is also plotted following Takagi et al\cite{Takagi}.
It is also shown the average low temperature gap $\Delta_d^{av}(p,T=0)$
that is proportional to the pseudogap temperature $T^*(p)$. }
\label{DoRnEJTc}
 \end{center}
\end{figure}

To show how $T_c(p)$ is calculated we plot also in Fig.(\ref{DoRnEJTc})  
the average low temperature  gap 
$\Delta_d^{av}(p,T=0)\approx \Delta_d^{av}(p,T_c)$
that enters into Eq.(\ref{EJ}).
As it is shown in Fig.(\ref{DoRnEJTc}), the results agree
with empirical formula $T_c(p)=95(1-82.6(p-0.16)^2)$. The 
dome-like shape is a consequence of the values of
$\Delta_d^{av}(p,T)$ that increases as $p$ decreases while $1/R_n(p)$
have opposite behavior, it is essentially zero near $p=0.05$ and
increases with $p$ (see $R_n(p)e^2/h$ plotted in Fig.(\ref{DoRnEJTc})).

In conclusion, the CH disordered free energy in cuprates
forms small electronic domains that induces local Cooper pairs and
also an array of Josephson junctions, as in a granular superconductor. 
This approach reproduces important experimental results and provides 
novel interpretations: The  two energy scales 
measured by electronic Raman  spectroscopy\cite{LeTacon}, tunneling
spectroscopy\cite{Krasnov} and  ARPES\cite{Damascelli,Lee}
are; one associated with the Josephson coupling energy
$E_J(p)$ that follows $T_c(p)$ and the other related with the 
d-wave mean superconducting gap $\Delta_d^{av}(p,T)$
that increases as the doping decreases.

Concerning the k-space anisotropy, the appearance of the Fermi arcs 
near the nodal direction is due to the lack of phase rigidity among 
superconducting domains combined with the small d-wave amplitudes 
around  the nodal directions. 
In this way the existence of Bogoliubov-like dispersions
above $T_c$, close to the antinodal directions, have the same
cause, namely the anisotropic d-wave geometry of the
intragrain superconductivity that is larger around these
directions and remains just at $T^*(p)$, i.e., above $T_c(p)$. 
Consequently the pseudogap phase is composed of these localized 
superconducting domains without phase coherence and with a nodal
metal structure.

I gratefully acknowledge partial financial aid from Brazilian
agencies FAPERJ and CNPq. 

\end{document}